%% file: 0.main.tex
\documentclass[sigconf,authorversion]{acmart}
\usepackage{subcaption}
\usepackage{colortbl}
\usepackage{multirow}
\usepackage{enumitem}
\setlist[enumerate]{leftmargin=*, itemindent=0pt}
\setlist[itemize]{leftmargin=*, itemindent=0pt}

\definecolor{lowcolor}{HTML}{D0F0C0} 
\definecolor{medcolor}{HTML}{FFFACD} 
\definecolor{highcolor}{HTML}{F08080} 

\newcommand{\riscv}{\mbox{RISC-V}}

\AtBeginDocument{%
  }

\setcopyright{acmlicensed}
\copyrightyear{2025}
\acmYear{2025}
\setcopyright{cc}
\setcctype{by}
\acmConference[CF '25]{22nd ACM International Conference on Computing Frontiers}{May 28--30, 2025}{Cagliari, Italy}
\acmBooktitle{22nd ACM International Conference on Computing Frontiers (CF '25), May 28--30, 2025, Cagliari, Italy}
\acmDOI{10.1145/3719276.3725186}
\acmISBN{979-8-4007-1528-0/2025/05}

\begin{document}

\title[Ramping Up Open-Source RISC-V Cores]{Ramping Up Open-Source RISC-V Cores: Assessing the Energy Efficiency of Superscalar, Out-of-Order Execution}
%
\author{Zexin Fu}
\orcid{0009-0004-6054-9078}
\affiliation{%
  \institution{ETH Zürich}
  \city{Zürich}
  \country{Switzerland}
}
\email{zexifu@iis.ee.ethz.ch}

\author{Riccardo Tedeschi}
\orcid{0009-0007-4483-9261}
\affiliation{%
  \institution{Università di Bologna}
  \city{Bologna}
  \country{Italy}}
\email{riccardo.tedeschi6@unibo.it}

\author{Gianmarco Ottavi}
\orcid{0000-0003-0041-7917}
\affiliation{%
  \institution{Università di Bologna}
  \city{Bologna}
  \country{Italy}}
\email{gianmarco.ottavi2@unibo.it}

\author{Nils Wistoff}
\orcid{0000-0002-8683-8060}
\affiliation{%
  \institution{ETH Zürich}
  \city{Zürich}
  \country{Switzerland}
}
\email{nwistoff@iis.ee.ethz.ch}

\author{César Fuguet}
\orcid{0000-0003-0656-2023}
\affiliation{%
  \institution{Univ. Grenoble Alpes, Inria, TIMA}
  \city{Grenoble}
  \country{France}}
\email{cesar.fuguet@univ-grenoble-alpes.fr}

\author{Davide Rossi}
\orcid{0000-0002-0651-5393}
\affiliation{%
  \institution{Università di Bologna}
  \city{Bologna}
  \country{Italy}}
\email{davide.rossi@unibo.it}

\author{Luca Benini}
\orcid{0000-0001-8068-3806}
\affiliation{%
  \institution{ETH Zürich}
  \city{Zürich}
  \country{Switzerland}
}
\affiliation{%
  \institution{Università di Bologna}
  \city{Bologna}
  \country{Italy}
}
\email{lbenini@iis.ee.ethz.ch}

\renewcommand{\shortauthors}{Z. Fu, R. Tedeschi, G. Ottavi, N. Wistoff, C. Fuguet, D. Rossi, L. Benini}

\begin{abstract}
Open-source RISC-V cores are increasingly demanded in domains like automotive and space, where achieving high instructions per cycle (IPC) through superscalar and out-of-order (OoO) execution is crucial. 
However, high-performance open-source RISC-V cores face adoption challenges: some (e.g. BOOM, Xiangshan) are developed in Chisel with limited support from industrial electronic design automation (EDA) tools. Others, like the XuanTie C910 core, use proprietary interfaces and protocols, including non-standard AXI protocol extensions, interrupts, and debug support.

In this work, we present a modified version of the OoO C910 core to achieve full RISC-V standard compliance in its debug, interrupt, and memory interfaces. 
We also introduce CVA6S+, an enhanced version of the dual-issue, industry-supported open-source CVA6 core.
CVA6S+ achieves 34.4\% performance improvement compared to the scalar configuration. 

We conduct a detailed performance, area, power, and energy analysis on the superscalar out-of-order C910, superscalar in-order CVA6S+ and vanilla, single-issue in-order CVA6, all implemented in GF22FDX technology and integrated into Cheshire, an open-source modular SoC platform. We examine the performance and efficiency of different microarchitectures using the same ISA, SoC, and implementation with identical technology, tools, and methodologies.
The area and performance rankings of CVA6, CVA6S+, and C910 follow expected trends: compared to the scalar CVA6, CVA6S+ shows an area increase of 6\% and an IPC improvement of 34.4\%, while C910 exhibits a 75\% increase in area and a 119.5\% improvement in IPC. However, efficiency analysis reveals that CVA6S+ leads in area efficiency (GOPS/mm²), while the C910 is highly competitive in energy efficiency (GOPS/W). This challenges the common belief that high performance in superscalar and out-of-order cores inherently comes at a significant cost in terms of area and energy efficiency.
\end{abstract}

%
%





\maketitle

\section{Introduction}
\label {sec:introduction}

\input{1.Introduction}

\section{Background}
\label{sec:backg}
\input{2.Background}

\section{Core Enhancements and Integration}
\label{sec:integration}
\input{3.2.Integration}

\section{Implementation}
\label{sec:impl}
\input{4.Implementation}

\section{Evaluation}
\label{sec:eval}

\input{5.Evaluation}

\section{Conclusion}
\label {sec:con}
\input{6.Conclusion}

\begin{acks}
This work has received funding from the Swiss State Secretariat for Education, Research, and Innovation (SERI) under the SwissChips initiative. This work was also supported in part through the ISOLDE (101112274) project that received funding from the HORIZON CHIPS-JU programme.
\end{acks}

\bibliographystyle{ACM-Reference-Format}
\bibliography{references}

\end{document}

%% file: 1.Introduction.tex
The recent trend toward increased autonomy in critical applications such as automotive and aerospace requires CPUs with robust, high-performance, and energy-efficient designs.
\riscv, an open-source instruction set architecture (ISA), has begun transitioning from its roots in low-power applications to more demanding, high-performance scenarios~\cite{henessy2019newgoldenage}. 
This shift necessitates the development of \riscv~cores that not only meet open ISA standards in terms of functionality, but also excel in instructions per cycle (IPC) through advanced execution strategies like superscalar and out-of-order (OoO). Moreover, open-source reference designs of these advanced cores are highly desirable, to enable auditing, fair benchmarking, and to democratize innovation.

While a number of closed-source superscalar, OoO RISC-V cores have been designed  (e.g. the products from Ventana \cite{ventana_veyron_v2}, Sifive \cite{sifive_p550_datasheet}, Semidynamics \cite{semidynamics_atrevido} or SOPHGO \cite{brown2023risc}) the landscape of open-source cores is much more scarcely populated.
Existing open-source \riscv~cores with superscalar and OoO capabilities, such as BOOM \cite{zhaosonicboom} and Xiangshan~\cite{xu2022towards}, have been developed using Chisel, a hardware description language (HDL) that, while powerful, is not widely adopted in the industry. This choice of language presents integration difficulties with standard electronic design automation (EDA) tools. 
On the other hand, XuanTie C910~\cite{chen2020xuantie}, an open-source high-performance \riscv~core, employs proprietary interfaces and protocols, including non-standard extensions to the AXI on-chip bus interface, as well as non-\riscv~compliant interrupt and debug support.
While these custom solutions may offer implementation advantages, they are a significant obstacle to widespread adoption.

As we push the performance boundaries of open-source OoO core in industrial practice, energy efficiency has become a critical factor, particularly for embedded systems, mobile devices, and data centers~\cite{shafique2014eda}. Power consumption of processors has become a limiting factor, leading to challenges in thermal management and battery life~\cite{brooks2007power}. This is especially relevant for \riscv~cores, which target applications ranging from low-power IoT devices to high-performance computing systems. 
However, conducting accurate and meaningful energy efficiency analyses for high-performance processors remains a major challenge.

Commercial cores are closed-source, making detailed comparative energy analysis very hard. 
Moreover, these cores are often implemented with proprietary technologies and methodologies, and integrated into complex platforms with numerous other closed source IPs (e.g., DDR controllers, PCIe interfaces) that can skew power consumption measurements. Different cores are typically implemented using different technologies, methodologies, platforms and efficiency analysis, making direct comparisons of their energy efficiency inaccurate and potentially misleading. The use of different benchmark suites or compilation tools and flags for different cores further complicates fair comparisons. These challenges have made it very hard to conduct an accurate, "apples-to-apples" energy efficiency comparison across different core microarchitectures, even when focusing only on \riscv~designs.

To address these challenges, 
we focused on three open-source cores with the same base ISA extensions: XuanTie C910~\cite{chen2020xuantie}, a high-performance, out-of-order, superscalar core which we modify and enhance for full \riscv~compliance; a dual-issue superscalar CVA6S~\cite{allart2024using} which we modify with several performance-enhancement features; and a single-issue vanilla CVA6~\cite{zaruba2019cost}. 
These three cores allow for a detailed, fair comparison, allowing us to explore the energy impact of out-of-order and superscalar execution.
All cores are ported to the same platform (Cheshire SoC~\cite{ottaviano2023cheshire}) and implemented using identical technology (GlobalFoundries 22 FDX) and methodologies to ensure that the core microarchitecture is the sole differentiating factor. 
Furthermore, we utilize the same benchmark binaries across all cores, compile with the same toolchain and optimization flags, to ensure consistent workload characteristics.

Our analysis covers metrics including performance (IPC), timing, area, power consumption, and energy efficiency. We perform a fine-grained area and power analysis, breaking down the consumption of the major microarchitectural components to identify specific key contributors to bottlenecks. This rigorous approach allows us to make meaningful comparisons between in-order and out-of-order cores. Our findings challenge the prevailing assumptions about their relative energy efficiency,  and provide valuable insights for future \riscv~core designs.
Our contributions are threefold:

\begin{enumerate}
    \item We present the first open-source, fully \riscv-compliant C910-based OoO core, featuring standard-conformant debug, interrupt, and memory interfaces.
    \item We present CVA6S+, the extended version of the superscalar variant of CVA6 \cite{allart2024using}, achieving a 34.4\% performance improvement over the baseline scalar CVA6 \cite{zaruba2019cost}.
    
    \item We conduct a detailed comparative analysis of performance, area, power, and energy efficiency across the three cores. This analysis uses the same base ISA, SoC environment, technology node (GlobalFoundries 22 FDX), tools, and methodologies, ensuring a fair and comprehensive comparison.
\end{enumerate}

Through extensive experimentation and analysis, we demonstrate that the area and performance rankings of CVA6, CVA6S+, and C910 align with expectations. 
Under the same L1 cache configuration, CVA6S+ and C910 incur 6\% and 75\% area overhead over the baseline CVA6, respectively. 
In terms of performance, CVA6S+ improves IPC by 34.4\%, while C910 achieves a 119.5\% gain.

While both configurations of the CVA6 core exhibit a small area footprint, CVA6S+ takes the lead in area efficiency (GOPS/mm²) and surpasses CVA6 in area-energy efficiency. The out-of-order C910 core is highly competitive in energy efficiency (GOPS/W), excelling not only in performance but also in achieving higher operating frequencies. 
These findings challenge the traditional assumption that high performance in superscalar and out-of-order cores inherently demands substantial trade-offs in area and energy efficiency~\cite{azizi2010energy, esmaeilzadeh2011dark, ronen2001coming}, offering valuable insights for designing and selecting processors in high-performance, energy-constrained applications.
The cores and SoCs described in this paper will be released to the public as open-source\footnote{https://github.com/pulp-platform/pulp-c910}.

This paper is organized as follows: Section~\ref{sec:backg} provides background on the cores and platforms under study. Section~\ref{sec:integration} details our methodology, including architecture enhancement and porting efforts made to the C910 and CVA6 cores. Section~\ref{sec:impl} presents our implementation flow and evaluation setup. Section~\ref{sec:eval} shows and discusses the results, and Section~\ref{sec:con} concludes the paper.

%% file: 2.Background.tex
\begin{figure*}[t]
    \vspace{-3mm}
    \centering
    \includegraphics[width=0.8\linewidth, height=10.5cm]{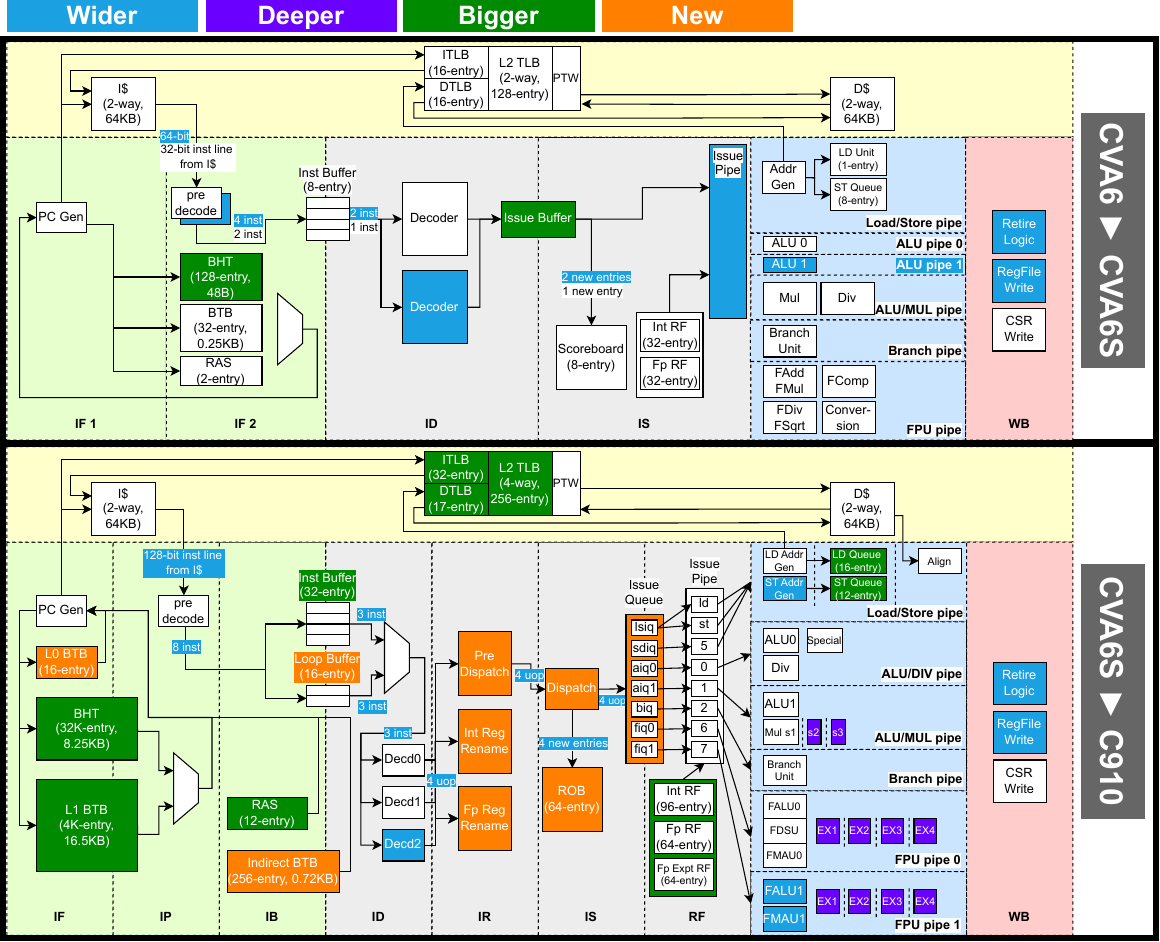}
    \caption{Architecture comparison of CVA6, CVA6S, and C910 cores}
    \label{fig:architecture_diff}
\end{figure*}

This section presents key prior works in the \riscv~ open-source ecosystem, which serve as the foundation for our enhancement, integration, and evaluation. Figure \ref{fig:architecture_diff} presents a detailed comparison of the architectures of the three cores evaluated in this work, emphasizing the design choices that result in varying tradeoffs among Instructions per Cycle (IPC), power consumption, and area metrics. Moreover, Table~\ref{tab:core_comparison} lists the microarchitecture parameters to provide a quantitative comparison at a glance.

\begin{table}[htbp]
\vspace{1pt}
\centering
\caption{CVA6, CVA6S, and C910 microarchitecture}
\label{tab:core_comparison}
\begin{tabular}{l@{\hskip 0.9mm}c@{\hskip 0.9mm}c@{\hskip 0.9mm}c}
\hline
\textbf{Feature} & \textbf{CVA6} & \textbf{CVA6S} & \textbf{C910} \\ \hline
Decode Width & 1 & 2 & 3 \\
Commit Width & 2 & 2 & 3 (up to 9) \\
Pipeline Stage & 6 & 6 & 12 \\
BHT Entries & 128 & 128 & 32K \\
BTB Entries & 32 & 32 & 4K \\
RAS Entries & 2 & 2 & 12 \\
Physical Regfile & 32 & 32 & 96 \\
FP Regfile & 32 & 32 & 64 \\
ROB Entries & 8 & 8 & 64 (up to 192) \\
ALU & 1 & 2 & 2 \\
MUL & 1 & 1 & 1 \\
DIV & 1 & 1 & 1 \\
BRU & 1 & 1 & 1 \\
FPU & 1 & 1 & 2 \\
Ld/St Width & 1 load + 1 store & 1 load + 1 store & 2 load/store \\
Load Queue Entries & 2 & 2 & 16 \\
Store Queue Entries & 4 & 4 & 12 \\ \hline
\end{tabular}
\vspace{-10pt}
\end{table}

\subsection{CVA6 and CVA6S (Superscalar CVA6)}
CVA6 is an application-class \riscv~CPU core developed as part of the PULP platform and now supported by OpenHW Group. It features a 64-bit in-order pipeline with six stages and supports both ASIC and FPGA implementations~\cite{zaruba2019cost}. CVA6 is parameterizable, allowing for configurable options such as 32- or 64-bit data path and address space, as well as optional floating-point support.

Its pipeline consists of six stages: two Instruction Fetch (IF) stages, followed by Instruction Decode (ID), Instruction Issue (IS), Instruction Execution (IE), and Writeback (WB). The IF logic can fetch either a single 32-bit instruction or two 16-bit compressed instructions per cycle from the instruction cache. Fetched instructions are placed into an 8-entry instruction buffer before entering the ID stage.
The IF stage also incorporates basic branch prediction structures to minimize pipeline stalls. 
These include a 128-entry Branch History Table (BHT) indexed by the lower bits of the program counter (PC), which uses 2-bit saturating counters to predict taken/not-taken outcomes. 
Additionally, a 32-entry Branch Target Buffer (BTB) caches target addresses for recently encountered branches, enabling rapid redirection of the fetch stream.

The IS stage has a single in-order issue port and an 8-entry scoreboard tracking the instruction commit order, which also functions as a Reorder Buffer (ROB). The execution stage includes dedicated pipelines for Load-Store, ALU, Multiplication and Division, Control Flow, and Floating Point (FP) operations. The WB stage can commit up to two instructions per cycle to accommodate the varying latencies of these pipelines.

Recently, Allart et al.\ introduced a superscalar configuration to CVA6~\cite{allart2024using}, resulting in a superscalar core referred to as CVA6S in this work. CVA6S enables dual-issue execution, achieving higher Instructions per Cycle (IPC) and making the core suitable for more demanding applications.
CVA6S retains the overall pipeline structure of CVA6, with modifications to support increased instruction throughput. The instruction fetch stage utilizes a 64-bit wide bus, allowing it to fetch up to two 32-bit instructions or four 16-bit compressed instructions per cycle. The decoding and issue logic are duplicated to enable dual-issue execution in an in-order configuration. The execution stage is enhanced with an additional ALU, which shares the writeback port with the Floating Point Unit (FPU). The FPU is however not used in their work due to the additional changes needed to properly identify structural hazards.

\subsection{XuanTie C910}
The XuanTie C910 is a high-performance, 12-stage out-of-order RV64GC core developed by Alibaba T-Head ~\cite{chen2020xuantie}. 
Compared to the scalar and superscalar CVA6 cores, the C910 architecture features a 12-stage pipeline, 3-issue width, and significantly larger microarchitectural structures, as detailed in Table~\ref{tab:core_comparison} and Figure~\ref{fig:architecture_diff}. 
These advancements establish the C910 as a high-performance, out-of-order processor. To address the heightened branch mispredict penalties resulting from its deeper pipeline, the C910 employs a sophisticated multi-level branch prediction framework. This includes a two-level Branch Target Buffer (BTB) for target resolution, comprising a 16-entry fully associative L0 BTB for zero-bubble prediction on hits and a 4K-entry set-associative L1 BTB for broader target coverage. 
The branch history table (BHT) utilizes a two-level adaptive mechanism that combines global and local history patterns for a higher prediction hit rate. It is also equipped with a deeper 12-entry Return Address Stack (RAS), ensuring precise function call/return handling. Additionally, a 16-entry Loop Buffer reduces fetch stalls during small-loop execution. Together, these mechanisms minimize pipeline flushes and optimize power efficiency.

C910 has a 64-entry reorder buffer (ROB) and deeper buffers in the load-store unit (LSU) to further improve its out-of-order capability, allowing it to hide more memory latency than CVA6 and CVA6S. It also has more functional units, including a dual-issue out-of-order LSU and another five-group execution cluster. 
Notably, the C910 core can compact up to three consecutive instructions into a single ROB entry and retire them simultaneously. This enables its ROB to theoretically hold up to 192 instructions and retire up to 9 instructions per cycle.
These characteristics position the C910 as a versatile core, achieving high throughput at the expense of increased area and power.
Despite its impressive capabilities, its proprietary interfaces for memory, interrupts, and debug system make it difficult to integrate into fully open-source ecosystems.

\subsection{Cheshire SoC}
Cheshire is a modular, Linux-capable SoC platform~\cite{ottaviano2023cheshire}. Designed to support application-class cores like CVA6, Cheshire provides a minimal, configurable host system with support for accelerators and manycore architectures. The SoC design emphasizes flexibility and scalability, making it a suitable candidate for experimenting with different \riscv~cores and extensions.

\begin{figure}[t]
    \vspace{-3mm}
    \centering
    \includegraphics[width=\linewidth]{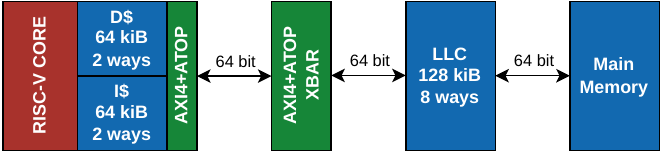}
    \caption{Memory hierarchy of the Cheshire SoC}
    \label{fig:cheshire_block}
    \vspace{-2mm}
\end{figure}

The memory hierarchy is depicted in Figure \ref{fig:cheshire_block}. It consists in two levels of caches in addition to the main memory. The \riscv~core includes both L1 Data and Instruction caches, which share a single AXI interface to interact as managers on the system crossbar. At the L2 level, a Last Level Cache (LLC) filters accesses to the main memory. All buses are 64 bits wide. The main memory used in the simulations is modeled as an ideal component with no internal latency. This work focuses on the \riscv~cores and their internal implementations rather than the higher level memory hierarchy. However, it is essential that all cores share the same memory hierarchy to ensure fair evaluations.

%% file: 3.2.Integration.tex
This section outlines the architectural modifications and system integration efforts focused on the XuanTie C910 and CVA6S core. Our contributions include standardizing the C910’s interfaces to comply with RISC-V standards and enhancing the CVA6S to improve its performance while maintaining energy efficiency.

Then, all the cores are integrated into the Cheshire SoC, enabling our comparative analysis on a unified platform. 
This consistent environment and the same benchmark binaries ensure our fair assessment of performance, power, and area trade-offs.

\subsection{Modifications to XuanTie C910}


The C910 core released by T-Head includes a tightly coupled internal SoC, 
but many signals are hardcoded, limiting modularity and scalability for integration with independently designed SoCs. 
Additionally, components like the interrupt and debug systems deviate from the RISC-V standard. 
To enable integration with the modular, open-source Cheshire SoC, we apply the following modifications:

\begin{itemize}
    \item \textbf{Interrupt Controllers:} We remove the internal CLINT and PLIC interrupt controllers from C910 and replace them with the RISC-V-compliant versions provided by Cheshire, improving standardization and compatibility.
    \item \textbf{Debug System:} C910's debug module and core debug support do not comply with the RISC-V debug specification. To ensure compatibility, we make the following modifications: \begin{itemize}
        \item Add RISC-V debug control and status registers (CSRs): \texttt{dcsr}, \texttt{dpc}, and \texttt{dscratch}.
        \item Introduced a debug request signal from the SoC and implemented a debug-mode exception in the decode stage, allowing instructions to trigger entry into debug mode.
        \item Enable the core to wake up from \texttt{wfi} (wait-for-interrupt) if the debug mode is triggered.
        \item Modify the pipeline control logic to ensure compatibility with the RISC-V debug module by: 
        1) updating debug-related CSRs, 
        2) redirecting the fetch stage to load the next program counter (PC) from the debug module, and 
        3) flushing the pipeline.
    \end{itemize}
    \item \textbf{Memory Interface:} We remove C910’s internal L2 cache, as Cheshire already includes an L2 cache that offers better flexibility. We also adapt the L1 cache interface, originally using a proprietary protocol derived from AXI-ACE with vendor-specific modifications. One such modification is a non-standard decrement mode for AXI burst transactions, which we convert into standard AXI increment bursts to ensure compatibility with the standard AXI/ACE protocol used in Cheshire.
\end{itemize}

After completing these modifications, the C910 core can be seamlessly integrated with the Cheshire SoC and boots Linux on a Xilinx VCU128 FPGA using the Cheshire software stack.




\subsection{Enhancements to CVA6S+}

We enhanced CVA6S with improvements targeting performance and execution hazard handling.

\begin{itemize}
    \item \textbf{Register Renaming:} We implemented register renaming to eliminate Write-After-Write (WAW) hazards by tracking the latest instruction writing to the Integer Register and Floating Point Register (FPR), ensuring correct operand forwarding when multiple inflight instructions target the same destination register.
    
    \item \textbf{Branch Prediction:} We implemented a two-level branch predictor with private history per entry~\cite{yeh1991two} (128 entries, 3-bit history), reducing misprediction penalties by 40\% and boosting performance by ~4.6\% on the Embench-IoT suite.

    \item \textbf{ALU-to-ALU Operand Forwarding:} Lightweight operand forwarding was implemented for scenarios where two ALU instructions are issued in the same cycle. This ensures that the source operands of the second instruction can directly depend on the result of the first, minimizing execution latency.

    \item \textbf{FPU Integration:} The Floating Point Unit (FPU), a feature missing on the upstream CVA6S, was fully integrated into the dual-issue architecture. This allows for the dual issuance of any floating-point operation alongside a non-floating-point operation. The only limitation is the combination of an FP store and another FPU operation, which requires additional FPR read bandwidth.
\end{itemize}

Together, these enhancements allow CVA6S+ to deliver higher throughput and better energy efficiency across both integer and floating-point workloads.

\subsection{L1 Cache}

Both CVA6 and CVA6S cores feature the HPDCache \cite{fuguet2023hpdcache}, while C910 has its own L1 cache design. 
The instruction cache is tightly coupled to the fetch unit, and the data cache to the load/store unit (LSU).

To ensure fairness in performance benchmarking, we configure the L1 caches of CVA6 and CVA6S to a size of 64 KB with two-way associativity for both the data and instruction caches to match the native configuration of C910. The C910 avoids aliasing issues in the data cache by employing a Physically Indexed, Physically Tagged (PIPT) scheme. In contrast, both CVA6 and CVA6s use a Virtually Indexed, Physically Tagged (VIPT) policy. This policy enables direct access to the cache line without going through the TLB only if the index bits are less or equal to 12 (corresponding to 4 KiB virtual memory pages). Unfortunately, a 2-way set associative cache violates the assumption and 3 bits of virtual address must be translated to fill the higher bits of the cacheline index.

Consequently, for CVA6 and CVA6S we implemented an additional antialiasing scheme that speculatively reuses the most significant bits from the last successful translation to populate the same bits of the index for subsequent transactions. These bits, which extend beyond the 12-bit virtual memory offset, require translation. In our solution, the request is aborted and retried if the translation reveals a mismatch between the predicted and translated bits; otherwise, it proceeds through the memory hierarchy as expected.


For backend timing evaluation, the critical path of all cores does not pass through the cache, so cache size does not affect timing. 


%% file: 4.Implementation.tex

This section details our implementation process. To ensure a fair and detailed comparison of the performance, timing, area, and power characteristics of the aforementioned three cores, we apply the same flow and methodology across all cores, minimizing any potential differences arising from the implementation and evaluation process.


\subsection{ASIC Flow}

We synthesize the designs using Synopsys Design Compiler 2022.12 and perform physical implementation with Cadence Innovus 20.12. 
All cores are implemented at the global worst-case corner (0.72 V, 125 $^{\circ}\mathrm{C}$) using standard cell and macro libraries of GF22 FDX technology.
These libraries support multi-threshold voltage (MTV) cells, which help balance power and performance. We apply a power recovery flow to improve power efficiency by replacing low-voltage threshold (LVT) cells with higher-threshold cells along paths with positive timing slack, while preserving timing closure.


We perform the power analysis using Synopsys PrimeTime 2022.03, based on switching activity from post-layout simulations with a common set of benchmark binaries.
Power is evaluated in a typical corner of 0.80 V at 25$^{\circ}\mathrm{C}$.
This consistent methodology enables a direct comparison of the power and energy consumption across the C910, CVA6S+, and vanilla CVA6 cores, offering new insights into the energy efficiency of out-of-order and superscalar execution.

\subsection{Benchmarking and Workloads}

We evaluate the performance of each core using the CoreMark \cite{gal2012exploring}, Embench-IoT suite~\cite{embench-iot} and the RaiderSTREAM suite \cite{beebe2022raiderstream}. The first two benchmarks are run with warm cache iterations, as the working set for all considered benchmarks fully fits within the L1 cache, resulting in negligible capacity and conflict miss rates. The latter one is instead used to evaluate each processor under sequential and non sequential memory access patterns with cold caches, resulting in several cache misses.
For performance benchmarking, we run full iterations of the test suite. For power analysis, we generate waveform VCD files for 30,000 cycles starting from the main benchmark body, following warm cache iterations where present.

%% file: 5.Evaluation.tex
\begin{figure*}[t]
    \vspace{-3mm}
    \centering
    \includegraphics[width=0.85\linewidth, height=5.2cm]{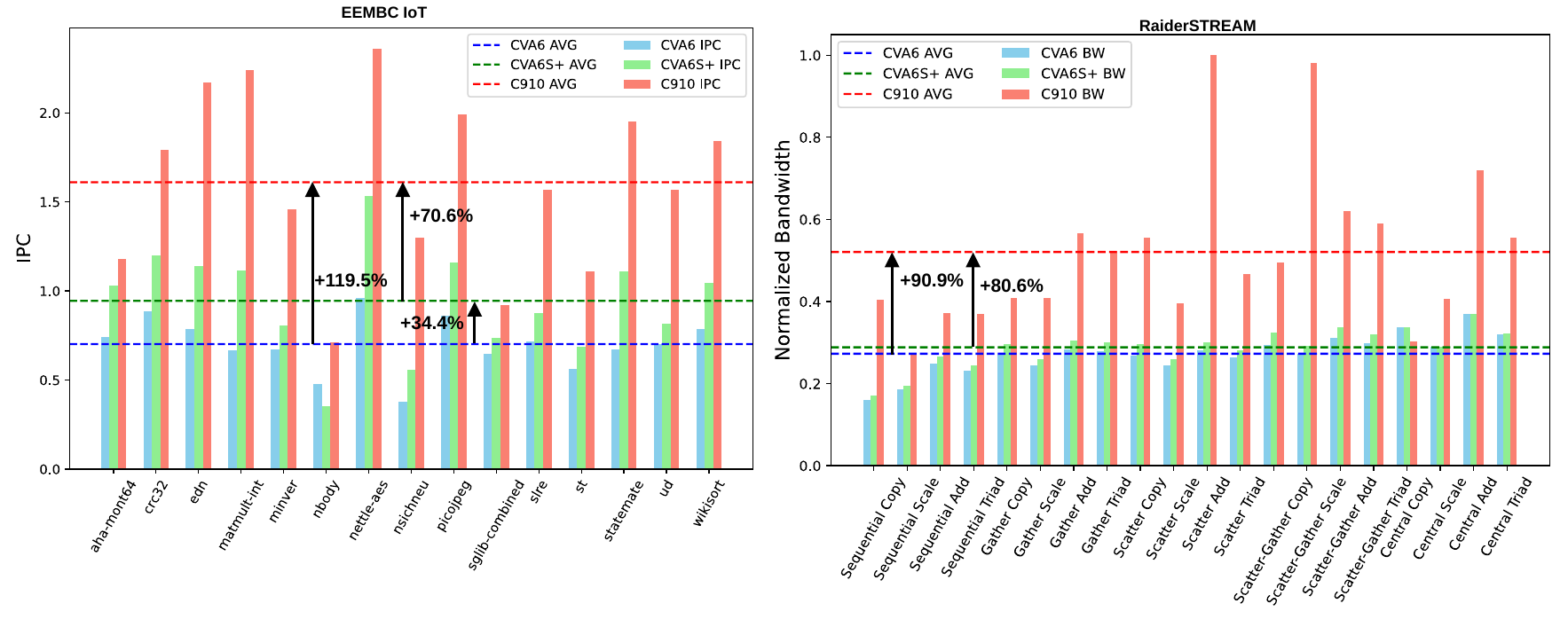}
    \vspace{-2.5mm}
    \caption{Performance comparison on Embench-IoT and RaiderSTREAM}
    \label{fig:ipc_graph}
    \vspace{-2mm}
\end{figure*}

\subsection{Performance}
\label{subsec:performance}


The performance of C910, CVA6S+, and CVA6 is assessed using instructions per cycle (IPC) for application benchmarks and normalized bandwidth for memory-intensive kernels.
Benchmarks from the Embench-IoT suite are selected to highlight various core attributes, including compute intensity, branch prediction accuracy, memory access patterns, and floating-point operations.
Embench-IoT is specifically designed for IoT devices, which typically operate with very limited on-chip memory. As a result, its memory access patterns are compact, enabling the L1 cache to hold all necessary data and significantly reducing cache misses. To provide a more comprehensive evaluation, the RaiderSTREAM benchmark is also employed. RaiderSTREAM is well-suited for analyzing both regular and irregular memory access patterns, placing substantial stress on the memory subsystem and effectively testing its performance under demanding conditions.

\textbf{C910} achieves the highest average IPC (1.61) and Coremark/MHz (4.86), due to its advanced out-of-order execution and 3-issue pipeline. 
Its multiple computing units (2 ALU, 2 FPU), and wide load/store unit (2 loads and stores) make it particularly effective in both compute-heavy and memory-bound tasks such as \textit{matmult-int} (IPC 2.24) and \textit{nettle-aes} (IPC 2.26). 
Additionally, the large branch prediction structures (BHT 32K, BTB 4K) enhance performance in branch-heavy workloads like \textit{statemate} (IPC 1.95).

Benchmarks with heavy double-precision floating-point operations, such as \textit{nbody} and \textit{st}, show lower IPC (0.71 and 1.11) due to long operation latencies and a limited floating-point pipeline relative to the core's issue width.

\textbf{CVA6S+} shows notable improvements over the scalar variant, achieving an average IPC of 0.94 and 2.84 Coremark/MHz, thanks to its superscalar execution.
It achieves a 66.5\% higher IPC in \textit{matmult-int} than the scalar configuration, 
enabled by the additional ALU and internal forwarding. Benchmarks like \textit{nsichneu}, 
which feature many nested \texttt{if} branches, also benefit from the two-level 
branch predictor, with IPC improving by approximately 46\% over base CVA6. 
However, the single FPU limits performance in floating-point-intensive workloads. 
For example, in \textit{nbody}, IPC drops to 0.35, performing worse than CVA6 (0.47). 
This is due to the additional ALU sharing the writeback port with the FPU, 
increasing structural hazards under floating-point-heavy conditions.


\textbf{CVA6} achieves the lowest average IPC (0.70) and Coremark/MHz (2.19) due to its single-issue, in-order pipeline and limited branch prediction.
Its single FPU further restricts performance in floating-point-heavy benchmarks, such as \textit{nbody} (IPC 0.47).

RaiderSTREAM benchmarks are instead evaluated through the normalized bandwidth over memory-intensive stream computations with regular and irregular patterns.
\textbf{C910} clearly shows its enhanced load/store capabilities, with the largest advantages observed on scatter and gather tests.
This performance edge stems from its out-of-order design, as well as much deeper ROB and LSU buffers, which enable it to issue a greater number of outstanding memory requests, effectively hiding more cache miss latency, and achieving higher overall performance.
In contrast, the \textbf{CVA6} and \textbf{CVA6S+} cores share the same LSU design, resulting in similar bandwidth outcomes for both.


\subsection{Timing}

\begin{figure}[t]
    \centering
    \includegraphics[width=0.8\linewidth, height=4.1cm]{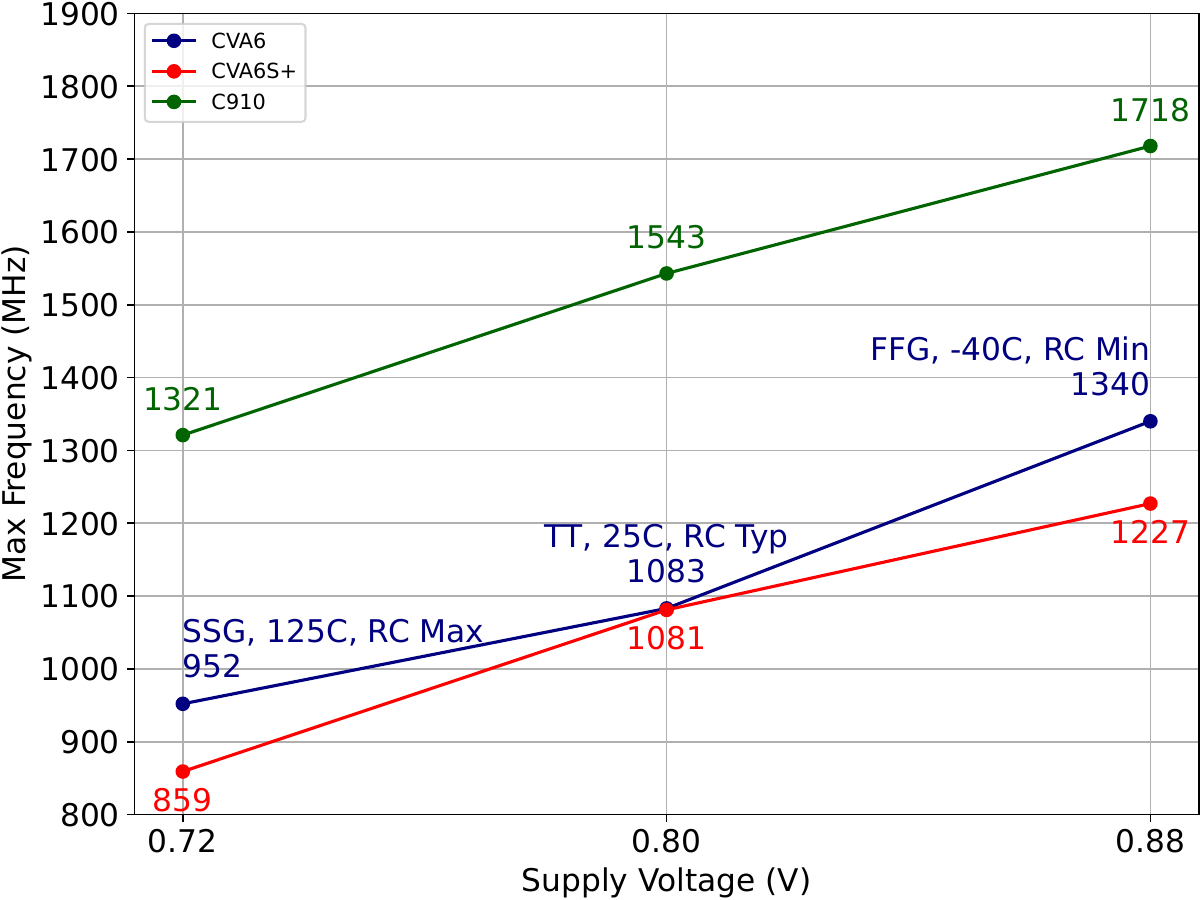}
    \caption{Maximum frequency at different supply voltages}
    \label{fig:max_frequency_vs_voltage}
    \vspace{-4mm}
\end{figure}

\begin{figure}[t]
    \centering
    \includegraphics[width=0.9\linewidth, height=4.1cm]{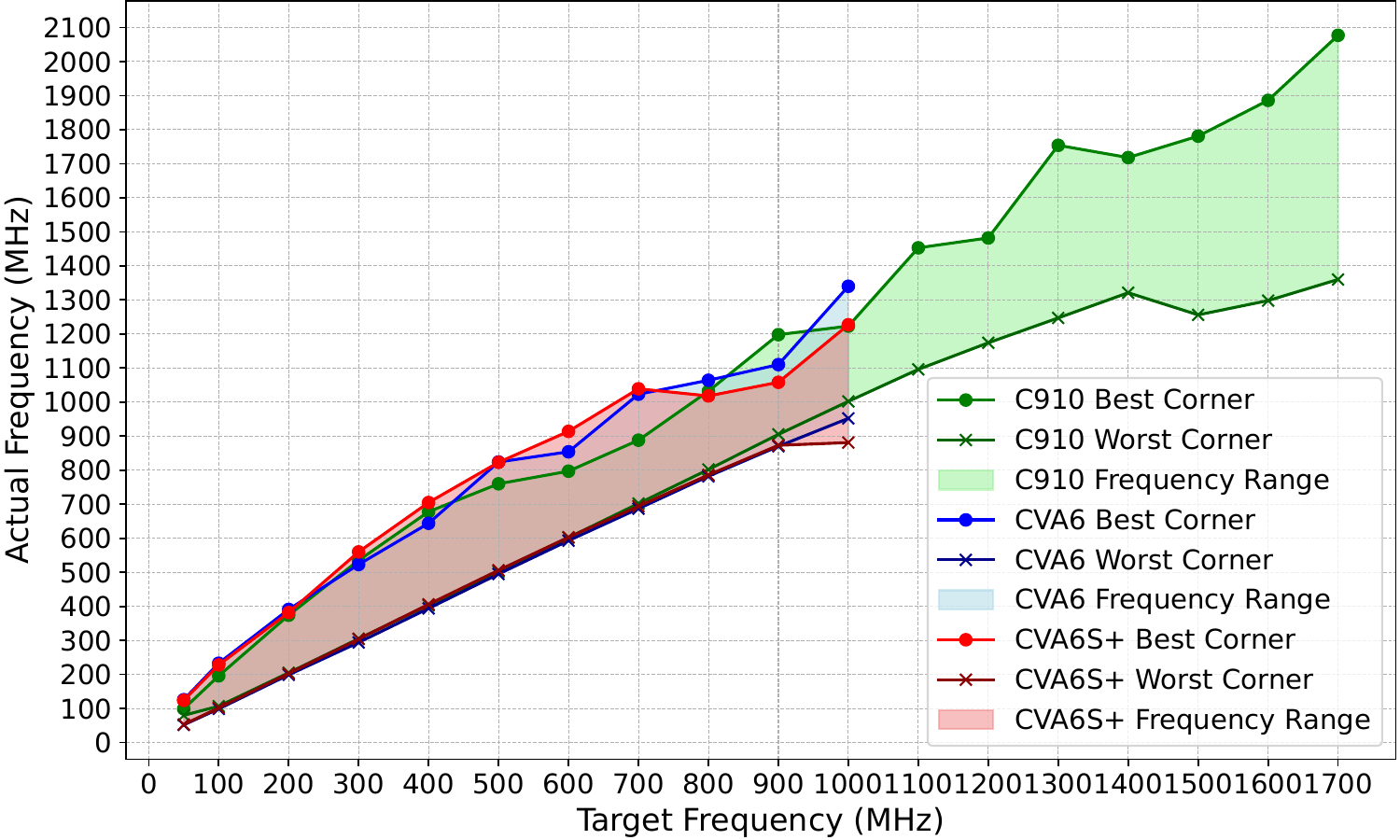}
    \caption{Frequency range of C910, CVA6S+, and CVA6 (best/worst corners)}
    \label{fig:actual_vs_target_freq}
    \vspace{-4mm}
\end{figure}

\begin{figure*}[t]
    \vspace{-3mm}
    \centering
    \includegraphics[width=0.95\linewidth, height=3.5cm]{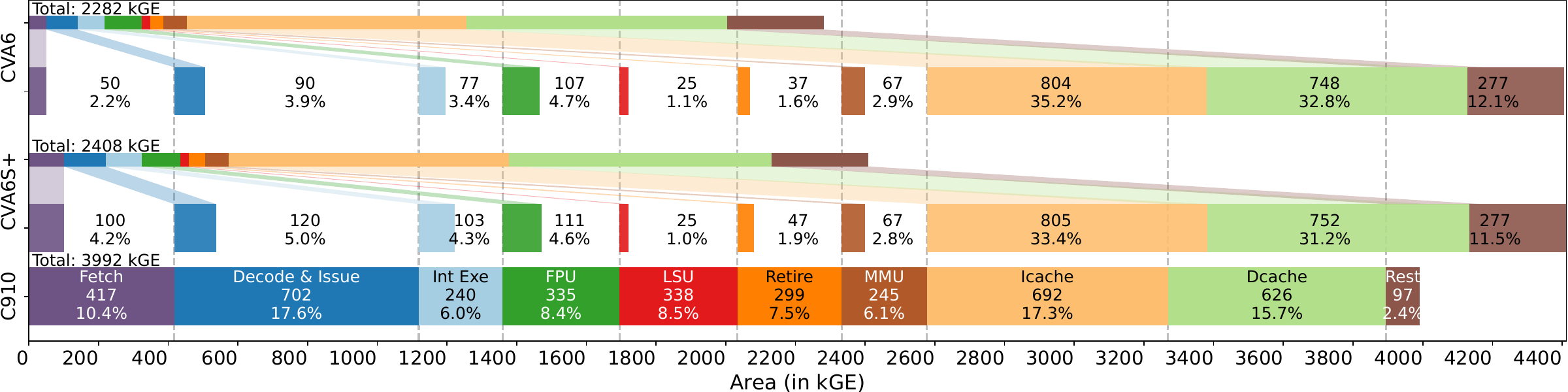}
    \caption{Area breakdown [kGE] near max target frequencies: 900\,MHz (CVA6, CVA6S+), 1300\,MHz (C910)}
    \label{fig:core_area_breakdown}
\end{figure*}

\begin{figure*}[t]
    \centering
    \includegraphics[width=0.95\linewidth, height=3.5cm]{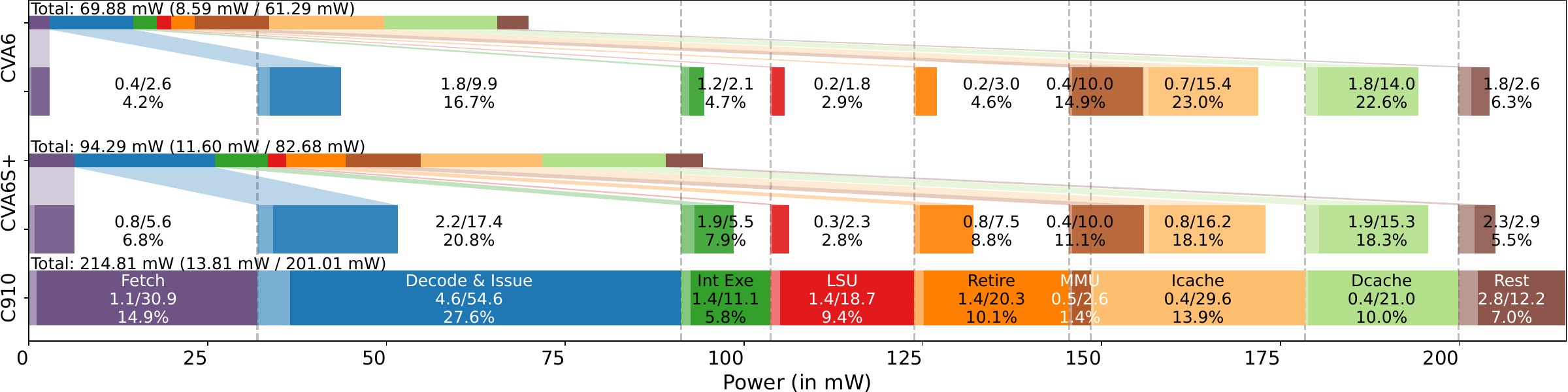}
    \caption{Power breakdown (leak/dyn) [mW] near max target frequencies: 900\,MHz (CVA6, CVA6S+), 1300\,MHz (C910)}
    \label{fig:core_power_breakdown}
\end{figure*}

The timing of C910, CVA6S+, and CVA6 is evaluated under best (FFG, $-40^{\circ}\mathrm{C}$, RCmin), typical (TT, $25^{\circ}\mathrm{C}$, RCtyp), and worst (SSG, $125^{\circ}\mathrm{C}$, RCmax) corners.
Timing is analyzed in terms of maximum achievable frequency.
Figure~\ref{fig:max_frequency_vs_voltage} compares the cores under similar timing optimization efforts, with LVT cell usage kept comparable: 58.7\% (CVA6), 61.9\% (CVA6S+), and 61.4\% (C910). 
Under the fast-fast corner (FFG, $-40^{\circ}\mathrm{C}$, RCmin) at 0.88\,V, 
\textbf{C910} reaches the highest frequency of 1718\,MHz, 
outperforming \textbf{CVA6} (1340\,MHz) and \textbf{CVA6S+} (1227\,MHz).
The higher frequency of C910 stems from its deeper 12-stage pipeline, enabling higher clock speeds despite a similar LVT cell usage to CVA6 and CVA6S+.
The CVA6 cores, with their 6-stage pipelines, are optimized for simpler, in-order execution, which constrains their maximum frequency despite comparable timing optimization efforts.
In Figure~\ref{fig:actual_vs_target_freq}, we can observe the maximum frequency range for each core, from the worst case to the best case process corners. C910 shows the potential for more frequency increases at higher voltages.

\subsection{Area}

The transition from CVA6 to CVA6S+ and the comparison with C910 highlight the differences in area utilization across key microarchitectural components. Figure~\ref{fig:core_area_breakdown} presents the area breakdown, measured in kilo gate equivalents (kGE), for each core operating near its respective maximum achievable frequency: 1300\,MHz for the C910 and 900\,MHz for both the CVA6 and CVA6S+.

\begin{figure*}[t]
    \centering
    \includegraphics[width=0.96\linewidth, height=2.8cm]{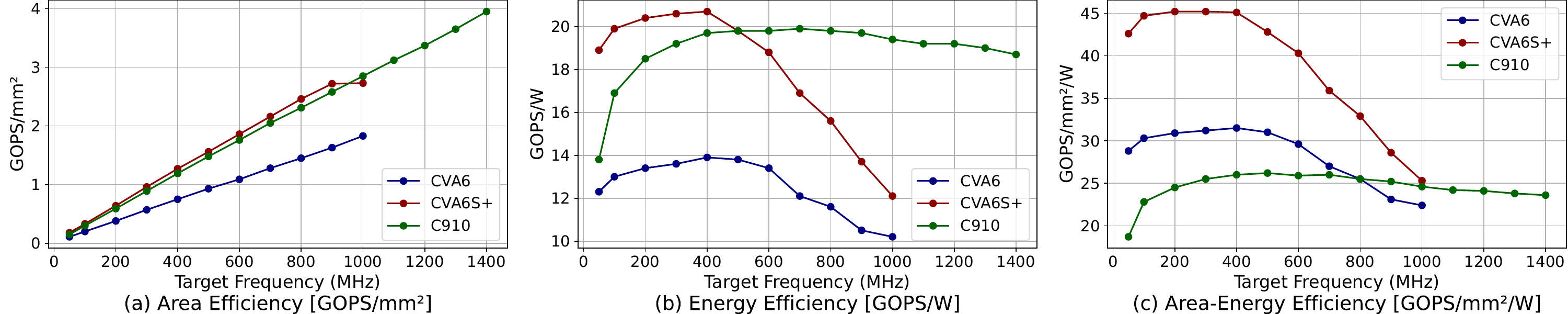}
    \caption{Area, energy and area-energy efficiency of the three cores under varying target frequencies}
    \label{fig:efficiency_multi_target}
\end{figure*}

\textbf{Transition from CVA6 to CVA6S+:} 
A direct comparison of the two cores can be made in Figure \ref{fig:core_area_breakdown} as they target the same frequency. 
Transitioning to CVA6S+ increases area by 16.5\% excluding caches, and 6\% including them.
Most of the increase comes from the Fetch, Decode, Issue, and Integer Execution units. 
Notably, the Fetch unit doubles from 50\,kGE in the scalar core to 100\,kGE in the superscalar version, primarily due to the redesigned branch predictor, which improves instruction throughput for dual-issue execution.
The Integer Execution unit area increases from 77\,kGE to 103\,kGE due to the addition of a second ALU and ALU-to-ALU operand forwarding, which improves execution efficiency but adds complexity to timing closure.
The Decode and Issue stages also grow from 90\,kGE to 120\,kGE for dual instruction decoding and issuing. Despite the area overhead, CVA6S+ has the best area efficiency, as shown in Section~\ref{sec:efficiency_eval}, with performance gains outweighing the cost.

\textbf{Compare CVA6 cores with C910:}
As shown in Table~\ref{tab:core_comparison}, the C910 features a wider and more complex microarchitecture compared to both the CVA6 and CVA6S+.
This is clearly illustrated in Figure~\ref{fig:core_area_breakdown}, which shows that, in the C910 core, the various stages and functional units occupy a larger proportion of the total area compared to caches, unlike in-order cores. The substantial area dedicated to Fetch, Decode, and Issue logic highlights the additional cost associated with OoO execution.
Although there is a notable performance gain, it is not as pronounced as the corresponding increase in area: 
as discussed in Section~\ref{subsec:performance}, the C910 delivers, on average, 2.2 times the IPC of CVA6 and 1.7 times the IPC of CVA6S+. At a target frequency of 900\,MHz, the area of the C910 is 1.75 times that of the CVA6 and 1.66 times that of the CVA6S+.


\subsection{Power}

\begin{figure}[t]
    \vspace{-3mm}
    \centering
    \includegraphics[width=0.96\columnwidth, height=4.5cm]{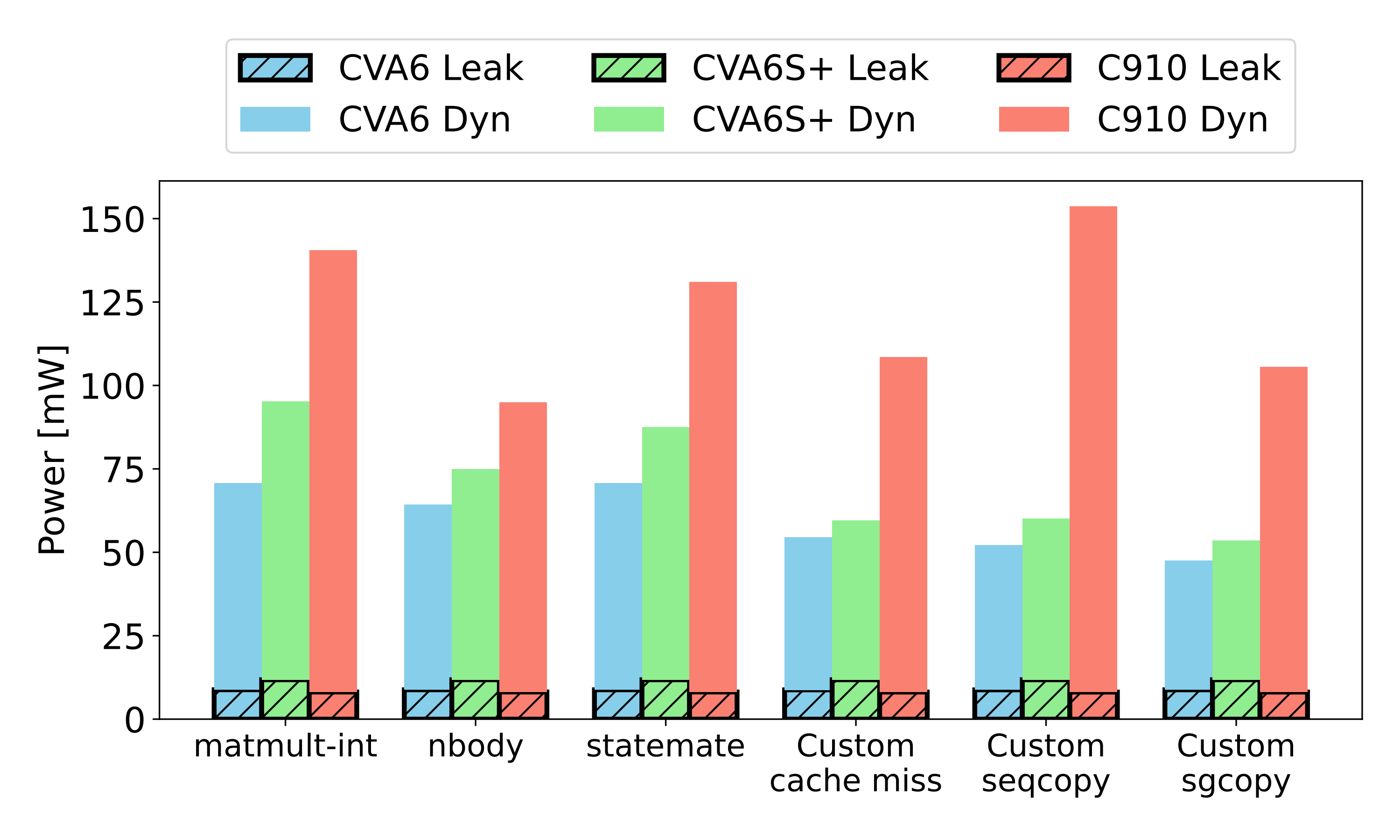}
    \vspace{-2mm}
    \caption{Power comparison on selected benchmarks at the same target frequency of 900\,MHz}
    \label{fig:power_results}
    \vspace{-2mm}
\end{figure}

The power consumption of each core is evaluated through post-layout netlist simulations, using a set of selected benchmarks. Embench \textit{matmult-int} (integer matrix multiplication), \textit{nbody}, and \textit{statemate} are utilized to represent Integer Compute-Intensive, Floating Point Compute-Intensive, and Branch-Intensive workloads, respectively. RaiderSTREAM (RS) \textit{seqcopy} and \textit{sgcopy} represent memory-intensive programs with regular and irregular access patterns. To facilitate post-layout simulation and power measurement, we implemented lightweight custom kernels that perform comparable operations with reduced runtime. The switching activity is extracted and analyzed using Synopsys PrimeTime 2022.03 (PT) under the typical corner (0.8V, TT, 25$^{\circ}\mathrm{C}$, RC typical).

Figure~\ref{fig:power_results} reports the total power, i.e. considering both dynamic and leakage power, for the different tests considering the same target frequency of 900\,MHz. As expected, the additional area in CVA6S+ translates into increased power consumption over CVA6.
This trend is also evident in the case of C910, particularly for memory-intensive kernels, where the power difference between C910 and the two in-order cores is substantial. This discrepancy primarily stems from C910's out-of-order and dual-issue LSU, combined with significantly larger buffers that accommodate multiple in-flight memory transactions. These features enable the C910 pipeline to maintain a higher level of utilization than the single-issue in-order LSU of the CVA6 cores.
It is worth noting that the C910 core exhibits lower leakage power, primarily because the 900\,MHz frequency represents a relaxed timing constraint for a deeply pipelined design. In contrast, the two in-order cores operate near their maximum achievable frequency, necessitating the use of LVT cells to boost performance. This optimization, however, comes at the cost of increased power consumption.

Figure~\ref{fig:core_power_breakdown} presents the detailed breakdown of the dynamic power density and leakage power for each core while executing the Embench-IoT \textit{matmult-int} benchmark near the respective maximum achievable frequency. Namely, C910 is evaluated at 1300\,MHz and CVA6 and CVA6S+ at 900\,MHz. The lighter colour represents leakage power, while the darker tone indicates dynamic power. 

When comparing the C910 to the two CVA6 implementations, it is evident that most of the additional power consumption arises from the units dedicated to fetching, decoding, issuing, and retiring instructions. These components contribute significantly to the increased design complexity when transitioning from an in-order to an OoO architecture. In the comparison between the CVA6S+ and CVA6, the additional power comes primarily from the architectural enhancements required to transform the core from single-issue to dual-issue. As can be seen in the figure, there is a sharp increase in the power draw of the units dedicated to fetching, decoding, and dispatching instructions. These changes cause the CVA6S+ core to draw 34\% more power compared to its single-issue counterpart.

\subsection{Area, Energy and Area-Energy Efficiency}
\label{sec:efficiency_eval}

We implement all three cores at multiple target frequencies to gain insights into the area, energy, and area-energy efficiency of these designs under different timing constraints. The power consumption of all cores is measured at the typical corner (0.8V, TT, 25$^{\circ}\mathrm{C}$, RCtyp), using VCD files generated from the post-layout netlist simulations of the \textit{matmult-int} benchmark.

\textbf{Area Efficiency},
measured in GOPS/mm\(^2\), allows us to understand how effectively each core utilizes its silicon area to deliver performance.
From Figure~\ref{fig:efficiency_multi_target}(a), we can see that CVA6S+ has the best area efficiency under the same timing constraint, while CVA6 has the worst area efficiency. As CVA6S+ originates from CVA6, its efficiency boost derives from the significant performance increase obtained with relatively minor area increase in the superscalar core with respect to the vanilla one. C910, despite its better performance, comes second to CVA6S+.
Across the three cores, the area efficiency improves as the target frequency increases. 
This trend reflects that, while achieving higher frequencies requires more area to meet timing constraints, the corresponding performance gains far outweigh the area increases. CVA6S+, with its dual-issue pipeline, sees the greatest gains in area efficiency, while the ability of C910 to reach much higher frequencies ensures it remains competitive despite a more complex and larger design.
The trend ends when the timing target becomes too tight, requiring more area to close timing, as seen at 900\,MHz for CVA6S+, where area efficiency plateaus.

\textbf{Energy Efficiency},
measured in GOPS/W, reflects the amount of energy consumed to perform a given workload.
The data in Figure~\ref{fig:efficiency_multi_target}(b) shows that as the timing constraint tightens, the energy efficiency of all cores initially increases and then begins to decline. This behavior is due to the different contributions of leakage and dynamic power at varying frequencies. At very low frequencies, leakage power consumes most of the energy, limiting efficiency gains. As frequency increases, dynamic power becomes the dominant factor, leading to improved energy efficiency.
However, at even higher frequencies, as the design approaches its timing upper limit, the EDA tool tends to insert high-energy-consumption cells to meet the timing constraints. This results in a decrease in energy efficiency.
This effect is particularly noticeable in CVA6S+, where energy efficiency drops at 500\,MHz and falls below that of C910. 
Once again, low IPC significantly penalizes CVA6 with respect to the other two cores.

\textbf{Area-Energy Efficiency},
measured in GOPS/mm\(^2\)/W, provides a combined perspective of how effectively each core balances area utilization and power consumption to deliver performance.
From Figure~\ref{fig:efficiency_multi_target}(c), we can observe that CVA6S+ achieves the highest efficiency figures. 
Although the XuanTie C910 delivers higher performance, its significantly larger area and power consumption at lower frequencies hinder its efficiency. As frequencies approach the 1GHz threshold, all cores converge to similar performance levels.

%% file: 6.Conclusion.tex
In this work, we present an in-depth analysis of \riscv~open-source cores, examining both scalar and superscalar (in-order) architectures, as well as out-of-order (OoO) machines. Our results, derived from implementing these cores on GF22FDX technology under various frequency constraints, highlight that simpler in-order cores maintain superior area efficiency compared to OoO cores. 
Specifically, the CVA6S+ core achieved an average of 1.65 times higher GOPS/mm², benefiting from a 34.4\% better IPC than CVA6, with a 6\% increase in area. This advantage extends to energy efficiency, particularly under low-frequency constraints. 
Overall, we observe that our enhanced CVA6S+ core outperforms or matches CVA6 in all efficiency metrics.
However, as the six-stage pipelines of CVA6 and CVA6S+ struggle to meet higher timing demands, the increased use of LVT cells degrades their overall efficiency.

Interestingly, our findings challenge the conventional belief that OoO cores are inherently much less energy-efficient than in-order cores~\cite{azizi2010energy, esmaeilzadeh2011dark, ronen2001coming}. We demonstrate that beyond the 500 MHz timing target, the C910 core surpasses in-order cores in energy efficiency.

These findings provide valuable guidance for selecting and designing processors in applications where both performance and energy efficiency are paramount. The in-order superscalar microarchitecture strikes an optimal balance in efficiency—encompassing area, performance, and overall effectiveness—due to its moderate increase in complexity compared to a single-issue in-order microarchitecture. Meanwhile, the superscalar OoO microarchitecture also demonstrates strong energy efficiency when high performance is the primary requirement.